\documentclass[a4paper,11pt]{article}
\usepackage{amsmath}
\usepackage{amssymb}
\usepackage{caption}
\usepackage{subcaption}
\usepackage{pos}

\usepackage[colorinlistoftodos,prependcaption,textsize=footnotesize]{todonotes}
\usepackage{xargs}

\newcommand{\e}{\mathrm{e}}
\newcommand{\xperp}{\mathbf x}
\newcommand{\yperp}{\mathbf y}
\newcommand{\vperp}{\mathbf v}
\newcommand{\kperp}{\mathbf k}

\newcommand{\sqrts}{\sqrt{s_\mathrm{NN}}}
\newcommand{\elrf}{\epsilon_\mathrm{LRF}}
\newcommand{\dd}{\mathrm d}
\newcommand{\tr}{\mathrm{Tr}}
\newcommand{\ii}{\mathtt{i}}
\newcommand{\xb}{\mathsf{x}}

\title{Limiting fragmentation in the dilute Glasma}

\author[a]{Andreas Ipp}
\author[a]{Markus Leuthner}
\author[a]{David I. M\"uller}
\author[b]{S\"oren Schlichting}
\author*[a]{Kayran Schmidt}
\author[c,d]{Pragya Singh}

\affiliation[a]{Institute for Theoretical Physics, TU Wien,\\
Wiedner Hauptstraße  8-10/136, A-1040 Vienna, Austria }

\affiliation[b]{Fakult\"at f\"ur Physik, Universit\"at Bielefeld,\\
D-33615 Bielefeld, Germany}

\affiliation[c]{Department of Physics, University of Jyv\"askyl\"a,\\
FI-40014 Jyv\"askyl\"a, Finland}

\affiliation[d]{Helsinki Institute of Physics, University of Helsinki,\\
FI-00014 Helsinki, Finland}

\emailAdd{ipp@hep.itp.tuwien.ac.at}
\emailAdd{mleuthner@hep.itp.tuwien.ac.at}
\emailAdd{dmueller@hep.itp.tuwien.ac.at}
\emailAdd{sschlichting@physik.uni-bielefeld.de}
\emailAdd{kschmidt@hep.itp.tuwien.ac.at}
\emailAdd{prasingh@jyu.fi}

\abstract{%
We discuss the local longitudinal scaling behavior of the dilute Glasma.
We gain insight into how the fragmentation region is dominated by the longitudinal structure of one of the two colliding nuclei in heavy-ion collisions and study the effect of the correlation scales of the nuclear color charge distributions.
We compare with results from the full dilute Glasma framework and analyze the rapidity limits of the fragmentation region.
From our findings, it follows that all scalar observables that can be constructed out of the dilute Glasma field strength tensor show limiting fragmentation behavior and the field strength tensor itself transforms locally via Lorentz boosts when changing the collision energy.
}

\FullConference{The XVIth Quark Confinement and the Hadron Spectrum Conference (QCHSC24)\\
 19-24 August, 2024\\
 Cairns Convention Centre, Cairns, Queensland, Australia\\}

\begin{document}

\renewcommand{\hookAfterAbstract}{%
\par\bigskip\bigskip
\textsc{ArXiv ePrint}:
\href{https://arxiv.org/abs/XXXXXX}{2502.XXXXX}
}
\maketitle

\section{Introduction}%
\label{intro}

The current understanding of relativistic heavy-ion collision experiments, such as at the Relativistic Heavy-Ion Collider (RHIC) or the Large Hadron Collider (LHC), is based on complex multi-stage models that describe the different physical processes valid for different timescales (see \cite{ALICE:2022wpn} for a recent review).
The description of the colliding nuclei before the collision is given by the Color Glass Condensate (CGC)~\cite{Gelis:2010nm,Gelis:2012ri}, an effective theory that predicts the Glasma~\cite{Lappi:2006fp} to appear as the solution of the classical Yang-Mills (YM) equations shortly after the collision.
The CGC requires nuclear structure models as input which are tested by analyzing the features discovered by experiments and comparing them to theory predictions.
One such feature is limiting fragmentation, first described in the CGC by \cite{Jalilian-Marian:2002yhb,Gelis:2006tb}, where rapidity profiles of different observables show universal behavior in the fragmentation region regardless of collider energy.
In these proceedings, we study the local longitudinal scaling
behavior of the three-dimensional, dilute Glasma~\cite{Ipp:2021lwz,Ipp:2024ykh,Leuthner:2025vsd} which encompasses non-trivial longitudinal structure and discuss the resulting limiting fragmentation phenomenon.
We use a fully three-dimensional nuclear model with parameters to control longitudinal and transverse correlation scales.


\section{Dilute Glasma field strength tensor}%
\label{sec:dilute-f}

We employ the CGC description for the highly relativistic, incoming nuclei labeled $A$ and $B$.
The valence partons inside the nuclei with a large momentum fraction $\xb$ are treated as classical color currents that source the highly occupied, small $\xb$ gluon fields.
The recoilless currents
$\mathcal J^\mu_{A/B}(x^\pm,\xperp) = \delta^\mu_\mp \rho_{A/B}(x^\pm,\xperp)$
are given in terms of the color charge distributions $\rho_{A/B}$ of each nucleus.
In light cone coordinates $x^\pm = (t\pm z)/\sqrt{2}$, only one vector component is nonzero and the $\rho_{A/B}$ only depend on one of $x^\pm$ and the transverse coordinates $\xperp$ (bold or with Latin indices $i,j,l$), where the $\pm$ is set by the direction of movement.
Solving the classical YM equations with the source term $\mathcal J^\mu_{A/B}$ in covariant gauge $\partial_\mu \mathcal A^\mu=0$ results in the single-nucleus gauge fields
\begin{align}
    \label{eq:A_poisson}
    \mathcal{A}^{\mp}_{A/B}(x^\pm,\xperp) = -(\nabla_\perp^2)^{-1} \rho_{A/B}(x^\pm,\xperp)= \phi_{A/B}(x^\pm,\xperp).
\end{align}

The dynamics of the gluonic field strength
$F^{\mu\nu} = \partial^\mu A^\nu - \partial^\nu A^\mu -ig \left[ A^\mu, A^\nu \right]$
in the forward light cone of the interaction region after the collision is governed by the classical YM equations
\begin{align}
    \label{eq:cym-full}
    D_\mu F^{\mu\nu}(x) =J^\nu(x), \qquad D_\mu J^\mu(x) = 0,
\end{align}
where
$D_\mu(\ldots) = \partial_\mu - ig\left[A_\mu,\ldots\right]$
is the gauge covariant derivative.
The dilute Glasma framework~\cite{Ipp:2021lwz,Ipp:2024ykh,Leuthner:2025vsd} starts with the ansatz
\begin{align}
    A^\mu(x) 
    = \mathcal A^\mu_A(x^+,\xperp) + \mathcal A^\mu_B(x^-,\xperp) + a^\mu(x), \quad 
    J^\mu(x) = \mathcal J^\mu_A(x^+,\xperp) + \mathcal J^\mu_B(x^-,\xperp)+ j^\mu(x). \label{eq:cyme-dilute}
\end{align}
The $a^\mu$ is the Glasma field of the full collision problem with the additional current $j^\mu$ that both are of the order $\rho_A^n \rho_B^m$ with $n,m\geq 1$ in the source terms.
To leading order, Eqs.~\eqref{eq:cym-full} are solved for $a^\mu$ in terms of $\phi_{A/B}$.
Then, the field strength tensor of the Glasma,
$f^{\mu\nu} = \partial^\mu a^\nu - \partial^\nu a^\mu$,
becomes abelian
\begin{align}
f^{+-}(x) &= -\frac{g}{2\pi} \int_{ \eta', \vperp} f_{abc} \, t^c \, \beta^{i,a}_A(x,  \eta', \vperp)  \,  \beta^{i,b}_B (x,  \eta', \vperp), \label{eq:f+-}
\end{align}
\begin{align}
f^{ij}(x) &= -\frac{g}{2\pi} \int_{ \eta', \vperp} f_{abc} \, t^c \Big( \beta^{i,a}_A(x,  \eta', \vperp) \,  \beta^{j,b}_B(x,  \eta', \vperp)  -\beta^{j,a}_A(x,  \eta', \vperp) \, \beta^{i,b}_B(x,  \eta', \vperp) \Big),\label{eq:fij}\\
f^{\pm i}(x) &=  \frac{g}{2\pi}  \int_{ \eta' ,\vperp} \frac{v^j}{|\vperp|} \frac{e^{\pm \eta'}}{\sqrt 2}  f_{abc} \, t^c \Big( \beta^{i,a}_A(x,  \eta', \vperp) \,  \beta^{j,b}_B(x,  \eta', \vperp)  -\beta^{j,a}_A(x,  \eta', \vperp) \, \beta^{i,b}_B(x,  \eta', \vperp) \nonumber \\
& \hspace{11em} \mp \delta^{ij}\beta^{l,a}_A(x,  \eta', \vperp)  \,  \beta^{l,b}_B (x,  \eta', \vperp)
\Big), \label{eq:f+-i}
\end{align}
where $\int_{ \eta', \vperp} = \iiint_{-\infty}^{\infty} \dd \eta' \dd^2\vperp$
and $t^c$ are the generators of SU($N_c$) with structure constants $f_{abc}$ and
\begin{align}\label{eq:beta}
    \beta^{i,a}_{A/B}(x,  \eta', \vperp) 
    &= \partial^i_{(x)}\phi^{a}_{A/B}(x^\pm\!- \frac{|\vperp|}{\sqrt{2}}e^{\pm  \eta'}, \xperp-\vperp),
\end{align}
where the $\partial^i_{(x)}$
are acting w.r.t.\ the $\xperp$.
The $\beta_{A/B}$ represent the only non-zero component of the field strength tensors of the nuclei $\mathcal{F}^{i\mp}_{A/B}=\partial^i\mathcal{A}^{\mp}_{A/B}=\beta^{i}_{A/B}$.

The integration in Eqs.~\eqref{eq:f+-}--\eqref{eq:f+-i} covers the backward light cone that is attached to the evaluation point $x=(x^+, x^-, \xperp)$ and is restricted to lightlike paths parametrized as
$v=(\frac{|\vperp|}{\sqrt{2}} e^{+\eta'},\frac{ |\vperp|}{\sqrt{2}} e^{-\eta'},\vperp)^\top$.
This geometric interpretation is illustrated in Fig.~\ref{subfig:backward-LC}, where for a given $\tau'=|\vperp|$, the integrand is only nonzero along the yellow highlighted curve inside the light gray interaction region where the color potentials of both nuclei contribute.

\subsection{Large $\eta_s$ approximation}%
\label{subsec:dilute-f-lf-approx}

\begin{figure}
    \centering
    \hfill\begin{subfigure}[b]{0.47\textwidth}
        \includegraphics[width=\textwidth]{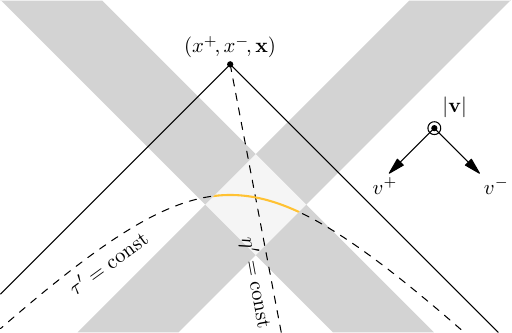}
        \caption{Evaluation close to $\eta_s=0$.}\label{subfig:backward-LC}
    \end{subfigure}\hfill\hspace{1em}\hfill
    \begin{subfigure}[b]{0.45\textwidth}
        \includegraphics[width=\textwidth]{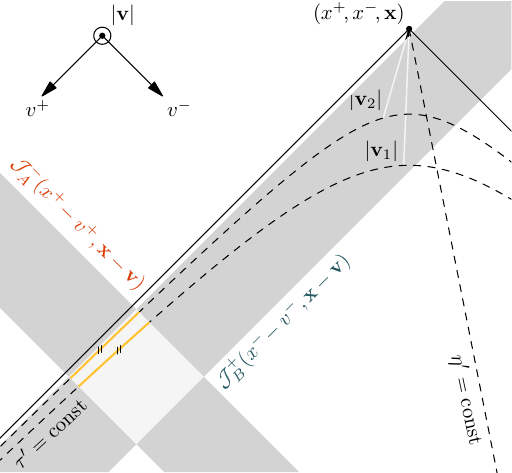}
        \caption{Evaluation at large $\eta_s$.}\label{subfig:backward-LC-large-eta}
    \end{subfigure}\hfill\ 
    \caption{(\subref{subfig:backward-LC}) Given the evaluation point $x = (x^+, x^-, \xperp)$, the contributions to the field strength tensor stem from yellow curves parametrized by $(\tau', \eta',\vperp)$ that are integrated over for each $\tau' = |\vperp|$. Adapted from~\cite{Ipp:2024ykh}. (\subref{subfig:backward-LC-large-eta}) Moving the evaluation point $x$ to large $\eta_s = \ln(x^+/x^-)/2$ deforms the yellow curves to nearly parallel lines cutting through $\mathcal{J}^+_B$ at a single value for $v^-$ at fixed $x^-$.}
    \label{fig:backward-LC-comp}%
\end{figure}

We study the behavior of the dilute Glasma field strength tensor in the fragmentation region where the spacetime rapidity $\eta_s = \ln(x^+/x^-)/2 \gg 1$.
As discussed in \cite{Ipp:2024ykh,Leuthner:2025vsd}, the evaluation points
$x^\pm = \e^{\pm \eta_s} \tau / \sqrt{2}$
are restricted to be outside of the nuclear tracks by shifting the origin of the Milne coordinate frame
along the $t$-axis.
Equations~\eqref{eq:f+-}--\eqref{eq:f+-i} are assembled out of products of the $\beta_{A/B}$ fields given by Eq.~\eqref{eq:beta}.
We rewrite these terms with explicit $\eta_s$ dependence under the $\eta'$ integral
\begin{align}\label{eq:beta-prod-milne}
    \int_{-\infty}^\infty \dd \eta'\ 
    \left[ \partial^i_{(x)}\phi^{a}_A(\frac{\e^{+\eta_s}}{\sqrt{2}}(\tau - |\vperp| \e^{\eta' -\eta_s}), \xperp-\vperp) \right] \left[ \partial^j_{(x)}\phi^{b}_B(\frac{\e^{-\eta_s}}{\sqrt{2}}(\tau - |\vperp| \e^{-\eta' +\eta_s}), \xperp-\vperp) \right],
\end{align}
and perform a shift $\eta' \rightarrow \eta'+\eta_s$ to isolate the $\eta_s$ dependence of the longitudinal arguments in a prefactor.
In the limit $\eta_s \gg 1$, the longitudinal argument of $\phi_A$ will only stay inside the track of nucleus $A$ for a very narrow $\eta'$ range because it is amplified by $\e^{+\eta_s}$.
Since the support of the integrand is limited to a narrow region around the point%
\footnote{%
Note, $\eta'=\bar{\eta}$ only contributes to the full $\eta'$ integral if
$\tau > \frac{\vperp^2}{\sqrt{2}\e^{\eta_s}u_\mathrm{max}}$,
where $u_\mathrm{max}>0$ is the longitudinal cutoff for the compact support of the nuclear fields in the longitudinal direction.
For large enough $\eta_s$ small values of $\tau$ are still allowed for the entire integration domain of $\vperp$.}
$\eta' = \bar{\eta} = \ln(\tau/|\vperp|)$, we can safely neglect the weak $\eta'$ dependence of $\phi_B$ and move it out of the integral.
We substitute
$u^+ = (\tau - |\vperp| \e^{\eta' -\eta_s}) \e^{+\eta_s} / \sqrt{2}$
to perform the remaining $\eta'$ integral over $\phi_A$ and obtain
\begin{align}
    \partial^j_{(x)}\phi_B^b(x^-\left(1-\frac{\vperp^2}{\tau^2}\right),\xperp-\vperp)\ \int_{-\infty}^{x^+} \dd u^+\ \frac{1}{x^+-u^+} \partial^i_{(x)}\phi^a_A(u^+,\xperp-\vperp)= \varsigma^j_B(x,\vperp)\ \varrho^i_A(x^+,\xperp-\vperp),
\end{align}
where we introduce
\begin{align}
    \varsigma^j_B(x,\vperp) &= \partial^j_{(x)} \phi_B( x^-\left(1 - \frac{\vperp^2}{\tau^2}\right), \xperp-\vperp), \label{eq:beta-lf} \\
    \varrho^i_A(x^+,\xperp-\vperp) &= \int_{-\infty}^{x^+} \dd u^+\ \frac{1}{x^+-u^+} \partial^i_{(x)}\phi_A(u^+,\xperp-\vperp). \label{eq:zeta-bb}
\end{align}
In the case of the $f^{\pm i}$ components, there is the additional factor $\e^{\pm \eta'}$ that leads to 
\begin{align}\label{eq:zeta-full}
    \varrho^i_A(x^+,\xperp-\vperp) \rightarrow \left(\frac{\sqrt{2}}{|\vperp|} \right)^{\pm 1} \int_{-\infty}^{x^+} \dd u^+\ \frac{(x^+ -u^+)^{\pm 1}}{x^+-u^+} \partial^i_{(x)}\phi_A(u^+,\xperp-\vperp).
\end{align}
The $\varrho_A$ is the effective field of nucleus $A$ that enters as a weighted longitudinal average.
We further simplify the weight factor by using that $x^+$ is far away from the track of nucleus $A$ s.t.\ $x^+ - u^+ \approx x^+ > 0$ for all $u^+$ where $\phi_A$ is nonzero.
Then, all the $x^+$ factors in Eq.~\eqref{eq:zeta-bb} as well as Eq.~\eqref{eq:zeta-full} can be pulled out of the $u^+$ integral and we define
\begin{align}
    \zeta^i_A(\xperp-\vperp) &= \partial^i_{(x)} \int_{u^+} \phi_A( u^+, \xperp-\vperp), \label{eq:zeta}
\end{align}
that only depends on transverse coordinates and appears in all the components of $f^{\mu\nu}$.
Equation~\eqref{eq:zeta} is the effective transverse field of nucleus $A$ as sourced by the transverse charge distribution $\rho_\perp(\xperp)=\int_{x^+} \rho_A(x^+,\xperp)$.
The physical picture in combination with the $\varsigma_B$ is close to a one-sided boost invariant limit where $\phi_B$ is cut into transverse slices for each fixed $|\vperp|$ and interacts with $\zeta_A$.
The only direct dependence on longitudinal nuclear structure comes from the different $\phi_B$ slices and is completely determined by nucleus $B$.

We obtain the field strength tensor of the dilute Glasma%
\footnote{The apparent divergence in Eq.~\eqref{eq:lf-fmunu-+i} for $\vperp^2\rightarrow 0$ is a coordinate artifact and vanishes when choosing polar coordinates for integrating the $\vperp$-plane.}
for the fragmentation region $\eta_s \gg 1$~\cite{Leuthner:2025vsd}
\begin{align}
    f^{+-}(x) &= -\frac{g}{2\pi} \frac{1}{x^+} \int_\vperp Y(x, \vperp), \label{eq:lf-fmunu-+-}  \qquad 
    f^{ij}(x) = -\frac{g}{2\pi} \frac{1}{x^+} \int_\vperp Y^{ij}(x, \vperp),
\end{align}
\begin{align}
    f^{+i}(x) &= \frac{g}{2\pi} \int_\vperp \left( Y^{ij}(x, \vperp) - \delta^{ij}Y(x, \vperp) \right) \frac{v^j}{\vperp^2}, \label{eq:lf-fmunu-+i} \\ 
    f^{-i}(x) &= \frac{g}{2\pi} \frac{1}{(x^+)^2} \int_\vperp \left( Y^{ij}(x, \vperp) + \delta^{ij}Y(x, \vperp) \right) \frac{v^j}{2}, \label{eq:lf-fmunu--i} \\
    Y(x, \vperp) &= f_{abc}t^a\ \zeta^{i,b}_A(\xperp-\vperp)\ \varsigma^{i,c}_B(x,\vperp), \label{eq:Y} \\
    Y^{ij}(x, \vperp) &= f_{abc}t^a \left(\zeta^{i,b}_A(\xperp-\vperp)\ \varsigma^{j,c}_B(x,\vperp) - \zeta^{j,b}_A(\xperp-\vperp)\ \varsigma^{i,c}_B(x,\vperp)\right). \label{eq:Yij}
\end{align}
The case for $\eta_s \ll -1$ can be worked out in full analogy, resulting in swapped roles of nucleus $A$ and $B$ and the light cone coordinates $x^\pm$.

In Fig.~\ref{subfig:backward-LC-large-eta} we compare the geometry of the limiting fragmentation (LF) approximation with evaluation point $x$ at large $\eta_s$ to $x$ at around mid-rapidity in Fig.~\ref{subfig:backward-LC}.
It is illustrative how the yellow curves deform to almost parallel lines that cut through $\mathcal J_B$ at an almost fixed $v^-$.
The weight factor in Eqs.~\eqref{eq:zeta-bb} and~\eqref{eq:zeta-full} is a correction for the yellow lines not being perfectly parallel to the track, but we argued that its effect is marginal and ignored it for Eq.~\eqref{eq:zeta}.

We emphasize that Eqs.~\eqref{eq:lf-fmunu-+-}--\eqref{eq:lf-fmunu--i} show local longitudinal scaling behavior (see Appx.~\ref{appx:demo} for full proof).
For scalar observables $\mathcal O(\eta_s)$ that are built out of the field strength tensor, this directly translates to
$\mathcal O_{\sqrt{s_1}}(\eta_s - Y_1) = \mathcal O_{\sqrt{s_2}}(\eta_s - Y_2)$,
that is: a change in collider energy $\sqrt{s_1} \rightarrow \sqrt{s_2}$ can be compensated by shifting with the difference in beam rapidity.
In particular, transverse integrated observables, like eccentricity moments $\varepsilon^n$ and azimuthal flow coefficients $\upsilon^n$, are predicted to show longitudinal scaling in the dilute Glasma.

\section{Nuclear model}%
\label{nucl-model}

The color charge distributions $\rho_{A/B}$ fluctuate event-by-event.
We assume the same generalized McLerran-Venugopalan (MV) model \cite{McLerran:1993ni,McLerran:1993ka} as in \cite{Ipp:2024ykh,Leuthner:2025vsd}, where the $\rho_{A/B}$ are sampled from a Gaussian probability functional determined by
$\langle \rho^a(x^\pm, \xperp) \rangle = 0$
and
\begin{align}
    \langle \rho^a(x^\pm,\xperp) \rho^b(y^\pm,\yperp)\rangle &= g^2\mu^2\delta^{ab}\sqrt{T(x^\pm, \xperp)}\sqrt{T(y^\pm, \yperp)} U_\xi(x^\pm-y^\pm)\delta^{(2)}(\xperp-\yperp), \label{eq:correlator_factorized}
\end{align}
with Yang-Mills coupling $g$ and MV scale $\mu$.
Equation~\eqref{eq:correlator_factorized} introduces three-dimensional profile functions which we choose to be Woods-Saxon distributions
\begin{align}
    T(x^\pm, \xperp) = \frac{c}{1+\exp(\frac{\sqrt{2(\gamma_\mathrm{beam} \, x^\pm)^2+\xperp^2}-R}{d})}, \qquad c = \frac{\gamma_\mathrm{beam}}{\sqrt{2}d\ln(1+\e^{R/d})},
\end{align}
with parameters $R$ and $d$ for nuclear radius and skin depth and longitudinal correlations of the shape
\begin{align}
    \label{eq:Uxi}
    U_\xi(x^\pm-y^\pm) = \frac{1}{\sqrt{2\pi \xi^2}} \exp(\frac{(x^\pm-y^\pm)^2}{4R^2/\gamma^2_\mathrm{beam}}) \exp(-\frac{(x^\pm - y^\pm)^2}{2\xi^2}),
\end{align}
with correlation length $\xi \leq 2 R / (\sqrt{2} \gamma_\mathrm{beam}$).
The Lorentz factor $\gamma_\mathrm{beam}$ is given by the collider energy and the other parameters are chosen as in \cite{Ipp:2024ykh}.
The resulting charge distributions are discretized on a lattice and their Fourier transforms in the transverse plane, $\tilde \rho$, are used to solve Eq.~\eqref{eq:A_poisson} via
\begin{align}
    \phi^a(x^\pm, \xperp) = \frac{1}{(2\pi)^2} \int_{\kperp}\frac{\tilde \rho^a (x^\pm, \kperp)}{\kperp^2+m^2}\e^{-\kperp^2/(2\Lambda_\mathrm{UV}^2)}\e^{-\ii\kperp\cdot\xperp},
\end{align}
where we introduce the ultraviolet regulator $\Lambda_\mathrm{UV} = 10$ GeV and infrared regulator $m$.
While $\Lambda_\mathrm{UV}$ is fixed to a value adequate for the chosen lattice spacing, different values of $m$ are used to control the size of transverse fluctuations of the nuclear color fields.
In the nonperturbative Glasma, this is the role of the saturation momentum $Q_s \propto g^2 \mu$, but in the dilute case, these physical parameters degenerate to overall prefactors~\cite{Ipp:2021lwz}.

\section{Numerical results}%
\label{results}

\begin{figure}
    \centering
    \begin{subfigure}[b]{0.47\textwidth}
        \includegraphics[width=\textwidth]{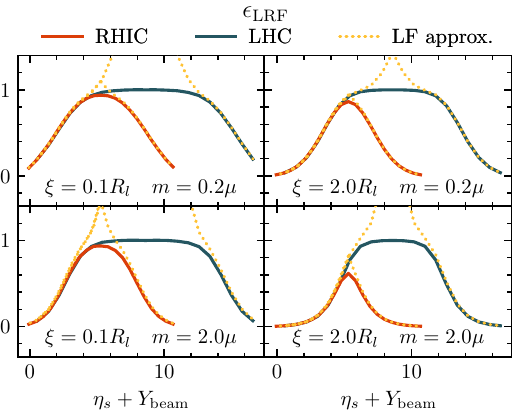}
        \caption{Local rest frame energy density $\elrf$.}\label{subfig:rap-prof-eps}
    \end{subfigure}\hfill
    \begin{subfigure}[b]{0.47\textwidth}
        \includegraphics[width=\textwidth]{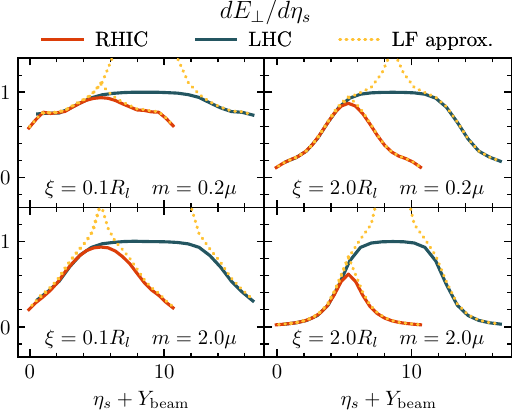}
        \caption{Transverse energy $dE_\perp/d\eta_s$.}\label{subfig:rap-prof-Ep}
    \end{subfigure}
    \caption{Comparison of the full dilute Glasma framework (solid lines) with the limiting fragmentation (LF) approximation (yellow). After shifting by the respective beam rapidities $Y_\mathrm{beam}=\mathrm{arcosh}(\gamma_\mathrm{beam})$ the flanks are perfectly matched between the two methods. For large longitudinal correlations $\xi$ the onset of the boost invariant plateau is recovered in the LF approximation. Data for a single event at $\tau=1.0$ fm/c with Au+Au at $\sqrts = 200$ GeV for RHIC (orange) and Pb+Pb $\sqrts=2700$ GeV for LHC (blue).}
    \label{fig:rap-prof}
\end{figure}

In Fig.~\ref{fig:rap-prof}, we compare numerical results for the LF approximation Eqs.~\eqref{eq:lf-fmunu-+-}--\eqref{eq:lf-fmunu--i} using the above nuclear model (yellow) to the results for the full dilute Glasma obtained in \cite{Ipp:2024ykh} (orange and blue lines) for typical RHIC and LHC setups.
The data are normalized to their value at $\eta_s=0$ and the $\eta_s$-axis is shifted by the respective beam rapidity $Y_\mathrm{beam} = \mathrm{arcosh(\gamma_\mathrm{beam})}$ for each collider energy.
Figure~\ref{subfig:rap-prof-eps} shows the local rest frame energy density $\elrf$ obtained by solving the Landau condition $T^\mu{}_\nu u^\nu = \elrf\, u^\mu$ with the energy-momentum tensor of the dilute Glasma, which to leading order in the sources can be calculated from the dilute Glasma field strength $f^{\mu\nu}$ s.t.\ $T^{\mu\nu} = 2\,\tr [f^{\mu\rho} f_{\rho}{}^{\nu} + \frac{1}{4} g^{\mu\nu} f^{\rho\sigma} f_{\rho\sigma}]$.
Figure~\ref{subfig:rap-prof-Ep} shows the sum of transverse pressures scaled with proper time $\tau$
\begin{align}\label{eq:Ep}
    \frac{\dd E_\perp}{\dd \eta_s} = \tau \int_\xperp\big(T^{xx}(\tau, \eta_s, \xperp)+ T^{yy}(\tau, \eta_s, \xperp)\big).
\end{align}
Both are integrated over a small region of the transverse plane, emphasizing the local nature of longitudinal scaling in the dilute Glasma.
The near-perfect agreement of the LF approximation to the full results is striking.
For small $\xi$, the LF approximation breaks down at smaller rapidities.
We explain this by noting that, in this case, the yellow curves in Fig.~\ref{subfig:backward-LC-large-eta} intersect different longitudinally coherent domains of nucleus $B$.
On the contrary, for large $\xi$, the yellow curves stay within a single coherent domain, and the LF approximation is valid up to the onset of the mid-rapidity plateau.

\section{Conclusion}%
\label{conclusion}

In the dilute Glasma framework the analytic expressions for the Glasma field strength tensor show local longitudinal scaling.
Using an approximation for the fragmentation region $\eta_s \gg 1$ we demonstrated this property and provided an interpretation for limiting fragmentation in terms of the longitudinal structure of only one of the colliding nuclei.
The current work is formulated in coordinate space.
A complementary manuscript studying the dilute Glasma in momentum space to provide a $k_\perp$-differential understanding of the dynamics is in preparation.
Additional studies could determine if local longitudinal scaling also holds for the non-perturbative Glasma.
Furthermore, the Glasma is only valid for the earliest stage of heavy-ion collisions and it is left to be studied how later state-of-the art descriptions, like relativistic hydrodynamics, change the longitudinal scaling behavior and dynamics predicted by the dilute Glasma.

\acknowledgments
{\footnotesize
\hyphenation{Deut-sche}
\hyphenation{For-schungs-ge-mein-schaft}
\hyphenation{Dok-to-rats-kol-leg}
ML, DM and KS are supported by the Austrian Science Fund FWF No.\ P34764\@.
SS acknowledges support by the Deutsche Forschungsgemeinschaft (DFG, German Research Foundation) through the CRC-TR 211 \lq Strong-interaction matter under extreme conditions\rq – project number 315477589 – TRR 211\@.
PS is supported by the Academy of Finland, by the Centre of Excellence in Quark Matter (project 346324), and European Research Council, ERC-2018-ADG-835105 YoctoLHC\@.
The computational results presented have been achieved in part using the Vienna Scientific Cluster (VSC).
}

\appendix
\section{Demonstration of local longitudinal scaling behavior}%
\label{appx:demo}

Starting from the solutions to the nuclear gauge fields in Eq.~\eqref{eq:A_poisson} at a given $\sqrts$, we look at a collision at a higher collision energy.
We introduce $w$, the relative rapidity that connects the rest frames of the nuclei at different $\sqrts$ via a Lorentz boost.
The nuclear fields at the higher collision energy are then given in terms of the lower energy fields as $\phi_{A/B}(x^\pm, \xperp) \rightarrow \e^{+ w} \phi_{A/B}(\e^{+ w} x^\pm, \xperp)$ 
which amounts to squeezing and scaling up the only non-zero component%
\footnote{This is not equal to boosting both fields with the same $w$, but to boosting each nuclei's field to higher energies taking into account their direction of movement.}.
This carries through to the $\varsigma$ and  $\zeta$ fields used in Eqs.~\eqref{eq:beta-lf} and~\eqref{eq:zeta}
\begin{align}
\varsigma^i_B(x^+, x^-, \xperp,\vperp) &\rightarrow  \partial^i_{(x)} \e^{+w} \phi_B( \e^{+w} x^-\left(1 - \frac{\vperp^2}{\tau^2}\right), \xperp-\vperp) = \e^{+w}\varsigma^i_B(\e^{-w}x^+,\e^{+w}x^-,\xperp, \vperp), \\
\zeta^i_A(\xperp-\vperp) &\rightarrow  \partial^i_{(x)} \int \dd (\e^{+w}u^+)\phi_A(\e^{+w}u^+, \xperp-\vperp) = \zeta^i_A(\xperp-\vperp).
\end{align}
The $\zeta$ field is unaffected by the modification because the factors can be absorbed by a change of variables under the integral.
Putting all together, the integrands for the components of the field strength tensor in Eqs.~\eqref{eq:Y} and \eqref{eq:Yij} each collect a factor of $\e^w$ and their $x^-$ ($x^+$) dependence is scaled by $\e^w$ ($\e^{-w}$)
\begin{align}
  Y(x, \vperp) & \rightarrow f_{abc}t^c\ \e^{+w}\ \zeta^{i,a}_A(\xperp-\vperp)\ \varsigma^{i,b}_B(\e^{-w}x^+,\e^{+w}x^-,\xperp,\vperp), \\
  Y^{ij}(x, \vperp) & \rightarrow  f_{abc}t^c\ \e^{+w} \left(\zeta^{i,a}_A\ \varsigma^{j,b}_B(\e^{-w}x^+,\e^{+w}x^-,\ldots) - \zeta^{j,a}_A\ \varsigma^{i,b}_B(\e^{-w}x^+,\e^{+w}x^-,\ldots)\right).
\end{align}
These factors can be rearranged to express the field strength tensor of the collision at higher energy as the boosted tensor of the collision at lower energy
\begin{align}
    f^{+-} & \rightarrow -\frac{g}{2\pi} \frac{1}{(\e^{-w}x^+)} \int_\vperp Y(\e^{-w}x^+,\e^{+w}x^-,\xperp,\vperp) =f^{+-}(\e^{-w} x^+, \e^{+w} x^-, \xperp), \\
    f^{ij} & \rightarrow -\frac{g}{2\pi}\frac{1}{(\e^{-w}x^+)} \int_\vperp Y^{ij}(\e^{-w}x^+,\e^{+w}x^-,\xperp,\vperp) = f^{ij}(\e^{-w}x^+, \e^{+w} x^-, \xperp), \\
    f^{+i} & \rightarrow \frac{g}{2\pi} \e^{+w}\int_\vperp \left( Y^{ij}(\ldots) - \delta^{ij}Y(\ldots) \right) \frac{v^j}{\vperp^2} = \e^{+w}f^{+i}(\e^{-w}x^+, \e^{+w} x^-, \xperp), \\
    f^{-i} & \rightarrow \frac{g}{2\pi} \frac{\e^{-w}}{(\e^{-w}x^+)^2} \int_\vperp \left( Y^{ij}(\ldots) + \delta^{ij}Y(\ldots) \right) \frac{v^j}{2}= \e^{-w} f^{-i}(\e^{-w}x^+, \e^{+w} x^-, \xperp),
\end{align}
such that finally $f^{\mu\nu} \rightarrow \Lambda^{\mu}{}_{\rho}(w) \Lambda^{\nu}{}_{\sigma}(w) f^{\rho\sigma}(\Lambda^{-1}(w) x)$ 
where $\Lambda(w)$ is the Lorentz boost operator with rapidity parameter $w$ acting on tensors and coordinates accordingly.
The field strength in the regime of longitudinal scaling for a collision at higher energies amounts to boosting the field strength of a lower energy collision with the relative rapidity $w$.

\bibliographystyle{JHEP}
\bibliography{references}

\end{document}